# Self-Propelled Agents and Group Social Force

**Peng Wang** **Peter Luh**

Brownian motion have long been studied on a diversity of fields, not only in physics of statistical mechanics, but also in biological models, finance and economic process, and social systems. In the past thirty years, there has been a growing interest in studying the model in self-propelled feature and interaction force such that the model also fits into study of social phenomenon of many individuals. This article will continue with this research trend and especially investigate the model in paradigms for a quantitative description of socio-economic process. We mainly discuss a class of collective decision process of Brownian agent/particles, where the stochastic process does not exist in the fluctuation in the traditional Brownian motion, but in selection among several discrete choices. Their decisions interacts with each other in a given social topology. To simplify our discussion the binary decision problem is particularly discussed where each agent only takes an alternative of two choices. Mathematically, we introduce a set of arrays to describe social relationship of agents in a quantitative manner, and the arrays deduce the group social force and opinion dynamics, which are useful to study complex social movement and self-organization phenomena including discrete-choice activities, social groups and de-individualization effect. Such agent-based simulation symbolizes a variety of collective activities in human society, especially in the field of economics and social science.

## I. RESEARCH BACKGROUND AND LITERATURE REVIEW

In the past thirty years, considerable efforts has been devoted in applications of physical paradigms for a quantitative description of social and economic processes, and a broad range of dynamical methods originally developed in a physical context have been applied to socio-economic phenomena. A substantial growth of research interest exists in dynamics of Brownian particles, which was initially discovered by British botanist, Robert Brown in 1827, and was later on mathematically described by Langevin Equation for non-equilibrium thermal process around 1908.

$$m_i \frac{d\boldsymbol{v}_i(t)}{dt} = \boldsymbol{F}^0(t) - \gamma\,\boldsymbol{v}_i(t) + \boldsymbol{\xi}_i \tag{1}$$

In statistical physics the stochastic effect is mainly discussed in the fluctuation term, which is commonly assumed to be a stochastic force with strength $D$ and $\delta$-correlated time dependence. In the case of thermal equilibrium systems we may assume that the fluctuation-dissipation theorem (Einstein relation) is applied. Traditionally in a physics sense, the Langevin equation is mainly used to describe collective motion of non-livling particles. When the equation is applied to modeling social movement of many living bodies, fluctuation force is not the focus of study, and it is necessary to introduce new concepts. Consequently, in the past thirty years many researchers were interested in advancing the model in self-propelled feature and interaction force such that the model also fits into study of social phenomena of many individuals.

---

Peng Wang previously studied in the Department of Electrical and Computer Engineering, University of Connecticut, Storrs, USA,

One group of researchers developed active Brownian particles, where particles that can store energy in an internal depot and use this energy to move actively and intentionally (Schweitzer et al., 1998; Ebeling and Schweitzer, 2001). The model is mainly used to describe biological unit with metabolism. The other major group refers to using social force or social field to describe interaction of many particles. Traditionally in statistical physics, the interactive forces are often omitted such as ideal gas, and researchers are more interested in the fluctuation force which represents the stochastic thermal effect. However, in the complex relations of socio-economic systems interaction force cannot be ignored, and it is crucial to have a system where the fluctuations due to unknown factors are not remarkably large compared to the deterministic interaction force. Such deterministic interaction enables us to describe herding effect and social group behavior, and it is named by social force in Helbing, Farkas, and Vicsek, 2000, and validation of such social interaction involves comparing the simulation of the model with associated observations drawn from real-world video-based analysis (Johansson et al., 2009).

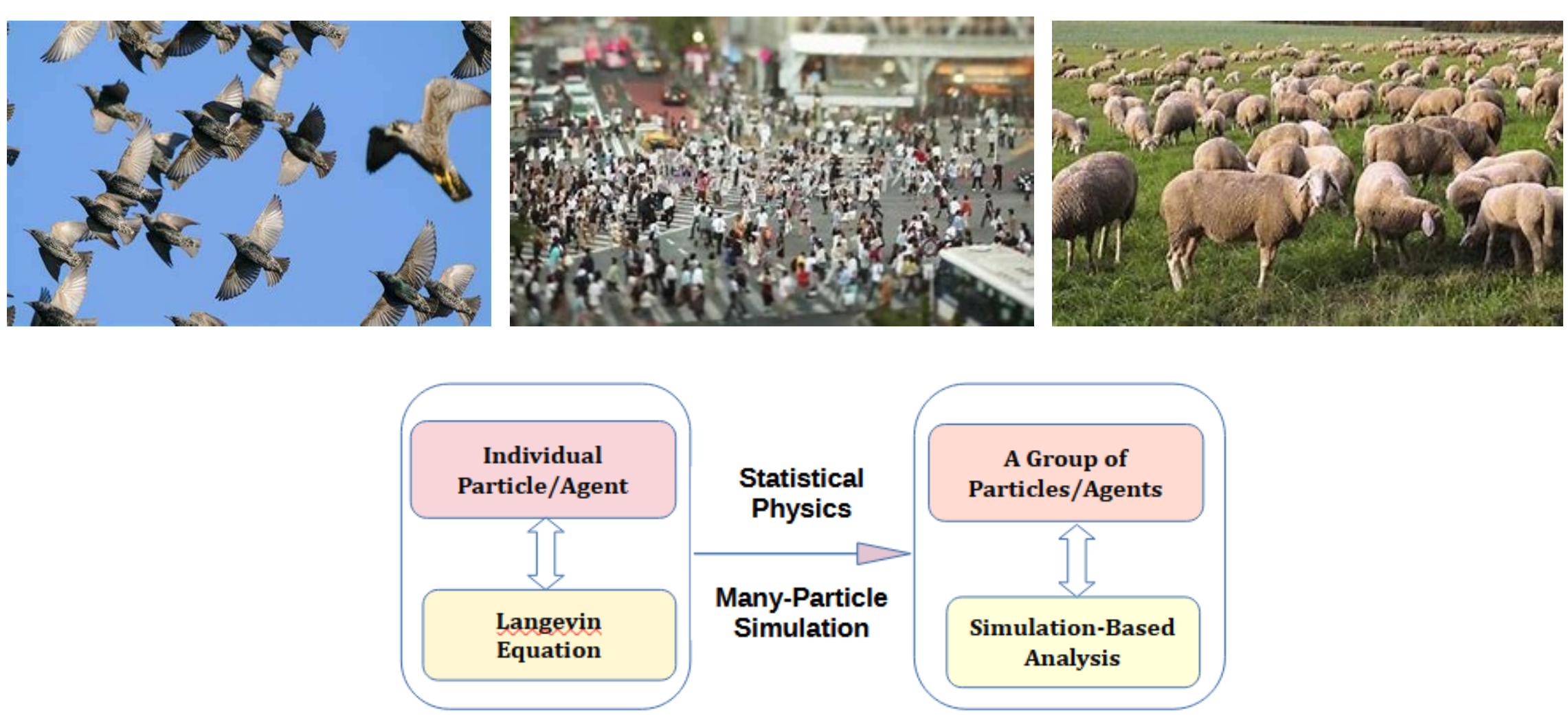

**Figure 1. From Particle Dynamics to Crowd Simulation**
**The Framework of Many-particle Simulation in Helbing, Farkas, and Vicsek, 2000**

More importantly, in social-economic process collective dynamics is widely studied, including social groups and de-individualization effect. A famous theory from socio-psychological perspective is called "automaton conformity" as depicted in Fromm, 1941, where individuals choose to de-individualize oneself to join a group. In order to model and simulate of this complex process, we should first assume two principles in our framework:

(i) We should admit that the stochastic effect does exist, but it seems not proper to simply ascribe such an effect to "thermal mechanics" and fluctuations. Rather, randomness in social behavior is largely due to selection among several deterministic known choices based on probability distributions. For example, a group of people are discussing whether the stock price will go arise or down, or voting for two candidates in a presidential election. Examples of interacting social agents may include actors in consumer market, voting behavior, or people in crowds or social animals in ecosystems. The opinion of each agent is described as the selection probability of limited alternatives, and probability distribution evolves as many agents interact in social context. Due to limited information available for each agent and bounded rationality it is well assumed that each agents' choice follow certain probability distribution. The resulting system is not limited to the theoretical analysis from the perspective of statistical physics, and thus we are inclined to use the general term of "agent" rather than "particle" throughout this article.

(ii) We need behavioral simulation jointly used with the above discrete choice process of opinion dynamics. The underlying reason is that even if people's preferences are well characterized by evolution of probability distributions, their behavior could be actually different from such opinions diversified at individual level because they may actively or passively hide their true preference in order to form a social group. In a psychological sense people may choose to defend or abandon their true freedom of choice in conscious mind or opinion, and de-individualization sometimes means "escape from freedom" (Fromm, 1941). This is the underlying mechanism of de-individualization in collective behavior, and force-based interaction is of critical importance to characterize this process in a collective sense. Such agents could be heterogeneous to some extent with a mixture of forced-based interaction and logic-based interaction.

In this article we will review and renew the above key concept of external field, and discuss prototypes of driving force and social force and in social-economic process. Our study refers to statistical physics, opinion dynamics control theory, traffic models as well as social and psychological principles. The rest of the paper is organized as below. Section 2 renews the definition of driving force to describe self-propelled agents, and opinion dynamics in social networks are introduced in Section 3. In Section 4 we will mainly present the group social force which is useful to combine individuals into groups. The concept of social groups is next established with a novel array-based structure in Section 5. In particular a class of two-state choice problem is especially discussed throughout the paper, in both theoretical and numerical approaches, and this problem is comparable to the classic binary system in statistical physics. A more complex opinion-behavior model with de-individualization effect is proposed in Section 6, and the conclusion remarks are presented in Section 7.

## II. Self-Propelled Agents

A conceptual difference between the self-propelled Brownian agents and traditional Brownian particles is that the driving force is not of external origin, but associated with each single particle and self-produced. As mentioned above, this implies each particle to have a kind of internal energy reservoir (Schweitzer et al., 1998; Ebeling et al., 1999). However, in this article we will mainly follow the idea of social force model. The social force model does not highlight how an individual gains energy from outside to realize motion. Rather, it emphasizes how an individual transforms the desirable motion in mind into the physical motion in reality. In other words, it is assumed that agents always has enough energy to achieve desirable motion, and the key issue exists in the desired state in one's opinion. The opinion is thus the essence of our study. As said "when there is a will, there is way," and the social force model emphasizes the importance of such subjective will, or as called desired state in one's opinion. The resulting self-driven agents are a paradigm for many active or living systems with conscious mind, where they are a simplified representation of the most fundamental dynamic behavior of living creatures such as cells, animals, and human pedestrian.

In order to describe self-propelled agents, sometimes Equation (1) is rewritten, where the external driving force $\boldsymbol{F}^0(t)$ is replaced by an individual driving force $k_i\boldsymbol{v}^0{}_i(t)$, implying that an agent is guided by a field which is inherently self-produced, but not generated from any external source.

$$m_i \frac{d\,\boldsymbol{v}_i(t)}{dt} = -k_i \boldsymbol{v}_i(t) + k_i \boldsymbol{v}_i^{\mathbf{0}}(t) + \boldsymbol{\xi}_i \tag{2}$$

In statistical physics $\boldsymbol{v}_i^0(t)$ is traditionally determined by a conservative potential field, and the resulting force component $k_i\boldsymbol{v}_i^0(t)$ exclusively depends on the position of the particles in the field. However, in this article the things become a little more complicated. First of all, each agent is guided by $\boldsymbol{v}^0{}_i(t)$ at individual level, and thus we add a subscript $i$ in notation of $\boldsymbol{v}^0{}_i(t)$ such that agents do not need to follow a common field as denoted by $\boldsymbol{F}^0(t)$ in Equation (1). In a sense, we assume that each agent has capability of memorizing a guiding field $\boldsymbol{v}_i^0(t)$ in one's opinion, and such opinion is actively selected in consistency of "self-driven" or "self-propelled" characteristics of agents. In other words, the guiding field is actively selected by agents among several alternatives, and thus it basically reflects opinions of agents. Here the convergence rate from $\boldsymbol{v}_i(t)$ to $\boldsymbol{v}_i^0(t)$ is also individualistic, as denoted by parameter $k_i$ in Equation (2). In Figure 2 we illustrate a typical example of binary choice problem, where two different guiding fields are demonstrated and agents are required to select one field to either move upwards or downwards to a destination location. Secondly, we assume that the self-produced field $\boldsymbol{v}_i^0(t)$ is basically time-dependent, and thus it is written as a function of time $t$. An agent thus could select to move upwards in this time slot while may change to move downwards during another time slot. This means that the opinion of each agent changes with time.

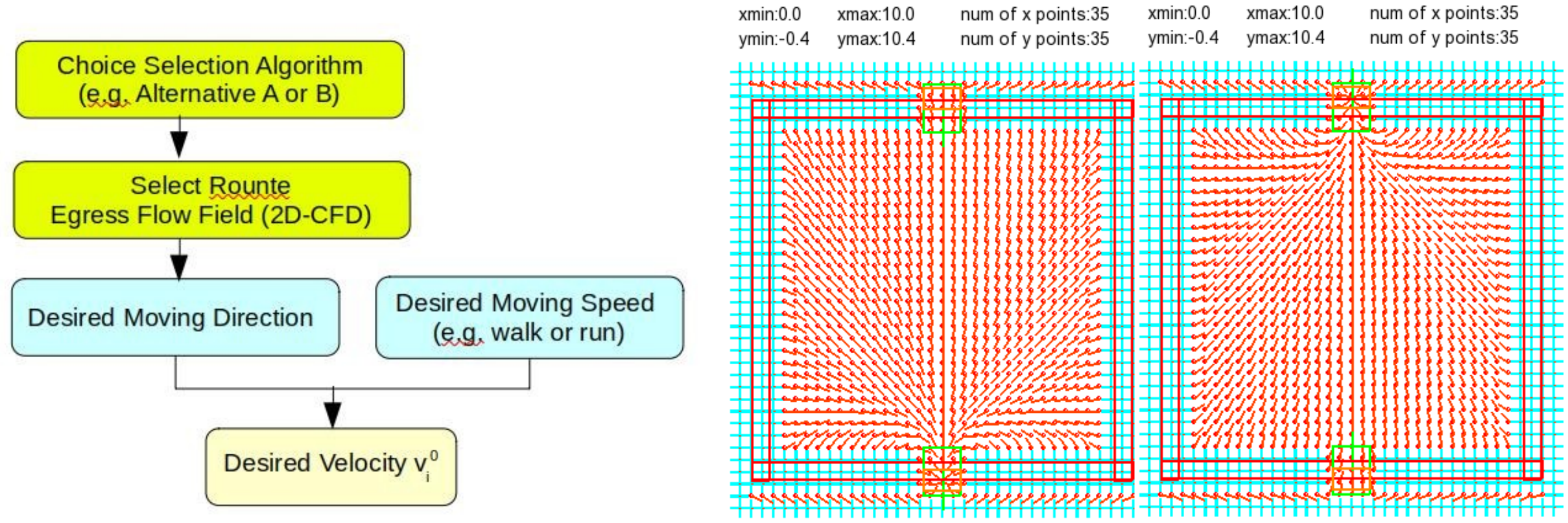

**Figure 2. Two Choices and Guiding Fields for Agents**

In a mathematical sense $\boldsymbol{v}_i^0$ is also rewritten as $\boldsymbol{v}_i^0(t) = v_i^0(t)\boldsymbol{e}_i^0(t)$, where $v_i^0(t)$ is the desired moving speed and $\boldsymbol{e}_i^0(t)$ is the desired moving direction. In a similar manner we also have $\boldsymbol{v}_i(t) = v_i(t)\boldsymbol{e}_i(t)$ where $v_i(t)$ and $\boldsymbol{e}_i(t)$ represent the physical moving speed and direction, respectively. In brief $\boldsymbol{v}_i^0(t)$ and $\boldsymbol{v}_i(t)$ are vectors with both directions and magnitudes. In social dynamics of many-particle systems, the magnitude and direction of $\boldsymbol{v}_i^0(t)$ are often treated as two separate issues as shown in Figure 2, and researchers are usually more interested in the direction $\boldsymbol{e}_i^0(t)$ such as in the Viscek flocking model (Viscek, 1995). In this article we will follow this widely-accepted idea such that $\boldsymbol{e}_i^0(t)$ is calculated by direction of a guiding field, and magnitudes $v_i^0(t)$ is kept constant. In other words, the magnitude of the guiding field is ignored in our computational model.

In practical computing each option of $\boldsymbol{e}_i^0(t)$ is calculated by a 2D conservative field that guides each individual agent to a destination location selected. In brief each candidate destination is considered as a sink point in a conservative potential field, and 2D Poisson Equation is formulated to calculate the road map towards the sink (Korhonen, 2017; Korhonen et. al., 2008). The flow solver is not presented in detail in this article, but we emphasize that this method is more suitable to describe behavior of living bodies on the background of social science and psychological theory. In a sense $\boldsymbol{v}_i^0(t) = v_i^0(t)\boldsymbol{e}_i^0(t)$ reflects a kind of cognitive map and social field referring to mind activities of creatures (Lewin, 1951, Helbing and Monlar, 1995, Helbing, 2001), and this is particularly useful to differentiate living bodies with conscious mind from non-living things like machines or robots.

Now we will discuss how to determine the guiding field $\boldsymbol{e}_i^0(t)$ by selection among several alternatives. Take the binary choice problem for example, where each individual agent is required to select one of two self-produced guiding field $\boldsymbol{e}_i^0(t)$. Suppose each individual agent is assigned with probability [$p1$, $p2$] to select either choice 1 or 2. Taking choice 1 or 2 means that an agent moves upwards or downwards, respectively. Here the probability distribution [p1, p2] could be initialized by the past selection frequency of two choices if certain historical data are available, or it can be simply given as [0.5, 0.5] if no bias is preset in agents' opinions. In the simulation process this probability distribution will be updated when agents interact with each other, and the probability measurement of [$p1$, $p2$] is changed in timeline and it forms a stochastic process to describe the preference of agents in selection of two alternatives. In the following discussion we will mainly use the term of probability. In a general sense of measure theory, the probability is a normalized real number that indicates agents' preference weight on each alternative in the choice set. The preference weight is equivalent to probability measurement because the more the agent prefers an option, the more likely he or she will choose it.

As a result, each individual agent is assumed to make a decision by selecting among several discrete alternatives, and decision is computed based on the continuous probability distribution, which critically reflects each individual's preference on the alternatives. The opinion of agents exist in the probability distribution and their decision and actions are computed from the random number generated from the probability distribution. The general idea is illustrated in Figure 3, where the continuous opinions (i.e., probability distribution) generate discrete decisions (i.e., random number), and further motivate agents' continuous action and movement in 2-dimensional space. Recall the above binary choice problem for example. There are two choices which could co-exist in an agent's opinion in a probabilistic sense. For example, one may prefer choice 1 with 0.7 probability, and choice 2 with 0.3 probability. However, when such an opinion is realized into behavior in the physical world, a decision should be reached by selecting one alternative definitely, and the decision is reached based on random number generated from the probability distribution. As a result, the decision will assign the corresponding guiding field to the agent and motivate the agent to move toward the destination location.

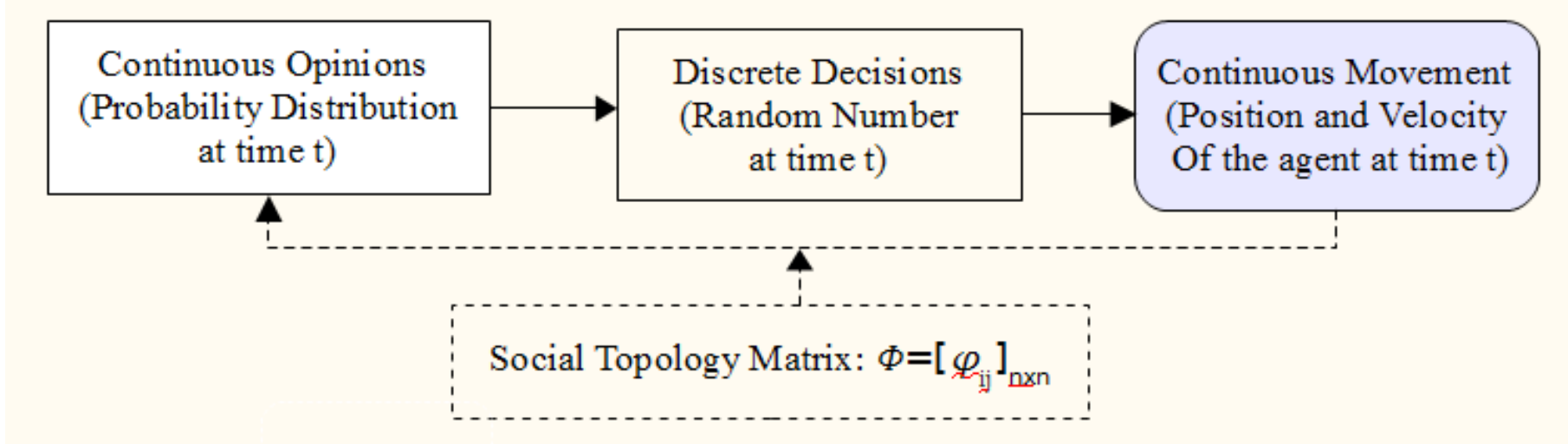

**Figure 3. Decision making process of selecting among discrete alternatives**

After $\boldsymbol{v}_i^0(t) = v_i^0(t)\boldsymbol{e}_i^0(t)$ is obtained from the guiding field, it is feasible to aggregate $k_i\boldsymbol{v}^0_i(t) - k_i\boldsymbol{v}_i(t)$ together as a linear form of driving force. Especially, in the social force model, the self-driving force is formulated in a linear feedback manner as $\boldsymbol{f}_i^{drv} = m_i(\boldsymbol{v}_i^0(t) - \boldsymbol{v}_i(t))/\tau_i$. By using the linear driving force, the actual velocity $\boldsymbol{v}_i(t)$ converges toward desired velocity $\boldsymbol{v}_i^0(t)$ in exponential rate. In the perspective of Langevin dynamics the driving term $\boldsymbol{v}_i^0(t)/\tau_i$ and friction term $-\boldsymbol{v}_i(t)/\tau_i$ together lead to an exponential adaptation of the velocity $\boldsymbol{v}_i(t)$ to the desired speed and direction of motion (Helbing, 2001; Helbing et. al., 2002). However, this convergence process may be disturbed by fluctuations $\xi_i(t)$. A major advantage of using Equation (3) is that we can differ so-called opinion and behavior for each agent. In specific $\boldsymbol{v}_i^0(t)$ is the target velocity existing in one's subjective opinion while $\boldsymbol{v}_i(t)$ is the physical velocity being achieved in the reality. Thus, $\boldsymbol{v}_i^0 - \boldsymbol{v}_i$ implies the difference between the subjective opinion and realistic situation, and it forms the driving force $\boldsymbol{f}_i^{drv}$ in a linear feedback manner.

Furthermore, the driving force $\boldsymbol{f}_i^{drv}$ in Equation (1) could be further generalized by $\boldsymbol{f}_i^{drv}=\boldsymbol{F}^{drv}(\boldsymbol{v}_i^0 - \boldsymbol{v}_i)$, where $\boldsymbol{F}^{drv}(.)$ is generally a monotonically increasing function such that $\boldsymbol{f}_i^{drv}$ increases with $\boldsymbol{v}_i^0 - \boldsymbol{v}_i$. This force describes an individual tries to move with a desired velocity $\boldsymbol{v}_i^0(t)$ and expects to adapt the actual velocity $\boldsymbol{v}_i(t)$ to the desired velocity $\boldsymbol{v}_i^0(t)$. Conceptually, the desired velocity $\boldsymbol{v}_i^0(t)$ is the target velocity existing in one's opinion while the actual velocity $\boldsymbol{v}_i(t)$ characterizes the physical speed and direction being achieved in the reality. The gap of $\boldsymbol{v}_i^0(t)$ and $\boldsymbol{v}_i(t)$ implies the difference between the human subjective wish and realistic situation, and it is scaled by a time parameter $\tau_i$ to generate the driving force. This force motivates one to either accelerate or decelerate such that the realistic velocity $\boldsymbol{v}_i(t)$ converge towards the desired velocity $\boldsymbol{v}_i^0(t)$. Based on control theory if we write $\boldsymbol{v}_i^{gap}(t) = \boldsymbol{v}_i^0(t) - \boldsymbol{v}_i(t)$ as an elementary term, it is feasible to also add its differential and integral term to Equation (3), and this is similar to construct a PID controller as below.

$$\boldsymbol{f}_i^{drv}=k_1\int \boldsymbol{v}_i^{gap}dt+k_2\boldsymbol{v}_i^{gap}+k_3\frac{d\,\boldsymbol{v}_i^{gap}}{dt}=k_1\int(\boldsymbol{v}_i^0(t)-\boldsymbol{v}_i(t))dt+k_2(\boldsymbol{v}_i^0(t)-\boldsymbol{v}_i(t))+k_3\frac{d(\boldsymbol{v}_i^0(t)-\boldsymbol{v}_i(t))}{dt} \tag{3}$$

The above equation exemplifies a general form of $\boldsymbol{f}_i^{drv}=\boldsymbol{F}^{drv}(\boldsymbol{v}_i^0 - \boldsymbol{v}_i)$. If $k_1$ and $k_3$ become both zero, Equation (3) is simply degenerated to Equation (2) where $k_2=m_i/\tau_i$. Similar to desired velocity $\boldsymbol{v}_i^0$, we will have another concept of the desired distance $d_{ij}^0$ which is the target distance in one's mind, specifying the distance that agent *i* desires to keep to agent *j*. The gap between the desired distance $d_{ij}^0$ and physical distance $d_{ij}$ further deduces the interaction force among agents. The interaction force term will be elaborated in Section 4. Before that we will introduce opinion dynamics on how individuals interact to form the $\boldsymbol{v}_i^0(t) = v_i^0(t)\boldsymbol{e}_i^0(t)$ based on the probability distribution of selecting several alternatives.

## III. On Decision Process of Many Agents

Now an important study topic is how to compute the continuous probability distribution, for example [p1, p2] in the binary choice problem, which refers to agent's opinion in timeline. This dynamical process characterizes how an individual aggregates the information acquired from other agents into his or her own opinion. Consider a group of agents among whom some process of opinion formation takes place. In general, an agent will neither completely follow nor strictly disregard the opinion of other agents, but will agree the opinions of others to a certain extent. We may partly interpret this milling process of many agents as a kind of herding instinct to form social groups, and it is a rooted nature in many specifies of social animals, helping individuals to gain a sense of safety, not only for human crowd, but also for herds, flocks and schools. In this section we will formulate opinion dynamics model and integrate it into the above framework of self-propelled agents.

Let $n$ be the number of agents under consideration, and the milling process of their opinions is mathematically described by different weights that any of the agents puts on the opinions of all the other agents. These weights are summarized compactly into a matrix $\boldsymbol{C}=[c_{ij}]_{n\times n}$ with $n$ rows and $n$ columns. The matrix thoroughly characterizes to what extent an agent will take the opinions of others in forming his or her own opinion. As mentioned before the opinion of an agent is expressed by a set of real numbers that represent probability distribution. Let $\mathrm{Prob}_{iq}(t)$ denote the probability that individual agent $i$ takes choice $q$ at time $t$, and the iterative opinion dynamics for agents $i$ is given by a linear combination of opinions of other agents.

$$Prob_{iq}(t+1)= \Sigma_j c_{ij}\cdot Prob_{iq}(t) \quad \text{with } \Sigma_j c_{ij}=1. \qquad OPIN(t+1)=C\cdot \mathrm{OPIN}(t) \qquad (4)$$

If opinions of $n$ agents are vertically stacked into a column vector as $\mathrm{OPIN}_q(t)=[Prob_{1q}(t), Prob_{2q}(t), \ldots Prob_{nq}(t)]^T$, and their opinion dynamics is compactly written as $\mathrm{OPIN}_q(t+1)=C\cdot \mathrm{OPIN}_q(t)$. Furthermore, if we horizontally stack $OPIN_q(t)$ for different choices into a row vector by $\mathrm{OPIN}(t) = (\mathrm{OPIN}_1(t), \mathrm{OPIN}_2(t), \ldots \mathrm{OPIN}_Q(t))$, and the above linear opinion dynamics is summarized in a matrix form of $OPIN(t+1)=C\cdot \mathrm{OPIN}(t)$. Suppose there are $Q$ alternatives in the choice set, OPIN(t) becomes a sort of "higher dimensional" opinions, which is $n$x$Q$ dimensional matrix. The i-th row of OPIN(t) is the complete probability distribution for agent $i$ to select among Q different choices.

There is also an interesting variation of this model developed by (Friedkin and Johnsen 1990, 1999). This model assumes that agent $i$ will always keep some prejudice to a certain extent $1-g_i$ and by a susceptibility of $g_i$ the agent is socially connected with other agents.

$$Prob_{iq}(t+1)= g_i\, \Sigma_j c_{ij}\cdot Prob_{iq}(t) + (1-g_i)\, u_{iq}(t) \quad \text{with } \Sigma_j c_{ij}=1. \qquad OPIN(t+1) = G\, C\cdot OPIN(t) + (I - G)\, u(t) \qquad (5)$$

Here $G$ is a diagonal matrix, namely, $G=\mathrm{diag}(g_1\ \ g_2\ \ldots g_n)$ with $0 \le g_i \le 1$, and $I$ is the identity matrix. The prejudice matrix $u(t)$ is also a probability distribution that desrcibes an agent's preference on two choices, and it is often assumed to be constant in the simulation process (Friedkin and Johnsen 1990, 1999). In a more general sense $u(t)$ is an input signal in a dynamical process, and it could be even considered as a noise signal which is randomly generated, characterizing a sort of random opinion in selection among alternatives. The resulting random signal $u(t)$ actually characterizes the fluctuation in the traditional Brownian particle of Equation (1), and it could be interpreted as certain amount of irrational opinion in social interaction. Consequently, the fluctuation effect in social opinion is represented by a "noise" distribution of $u(t)$ in selection among several deterministic known choices.

Now the opinion of each agent is updated by a linear combination of other agents' opinions, and the core of our discussion exists in matrix $\boldsymbol{C}=[c_{ij}]_{n\times n}$, where $c_{ij} \ge 0$ is the social weight that agent $i$ gives to agent $j$. In specific $\Sigma_j c_{ij}=1$ means that $\boldsymbol{C}=[c_{ij}]_{n\times n}$ is a stochastic matrix, i.e., a non-negative matrix with all its rows summing up to 1. The existing theory in linear algebra suggests that opinion OPIN(t) converges if matrix $C$ satisfies certain conditions. For example, if $C$ is irreducible, consensus of all $n$ agents requires that $\boldsymbol{C}=[c_{ij}]_{n\times n}$ is primitive. For the classical case of constant weights and enough confidence among agents a typical phenomenon is consensus (French 1956, DeGroot 1974, Lehrer 1975). A more complicated case is for time-variant $\boldsymbol{C}(t)=[c_{ij}(t)]_{n\times n}$, where $c_{ij} \ge 0$ is dynamically changing as time proceeds. This situation implies that the connectivity of the graph is updated as agents move and interact, and the social topology among agents thus become time-variant.

If we interpret the matrix $\boldsymbol{C}=[c_{ij}]_{n\times n}$ by graph theory such that $c_{ij} > 0$ represents an arc directed from agent $j$ to agent $i$, then matrix $\boldsymbol{C}=[c_{ij}]_{n\times n}$ actually specifies a directed graph $G(\boldsymbol{C})$ that describe the social topology of $n$ agents. Take Figure 4 for example and we have three agents and their social connection is illustrated by a directed graph. The directed arc from individual 1 to 2 means that individual 2's opinion is impacted by individual 1 with a given social weight of 0.9. The self-loop in the graph is is also indexed by a social weight, describing how an agent will keep his own opinion in the iterative process of opinion dynamics, and it is equal to one minus the sum of all the weights on the input arc towards the agent. The weight on the self-loop corresponds to the diagonal element in $c_{ii}$ in matrix $\boldsymbol{C}$, and in this article we will assume that these weights could be given very small, but cannot be zero, namely $c_{ii}>0$, meaning that an agent must hold part of his own opinion when interacting with others and forming new opinion in next time step.

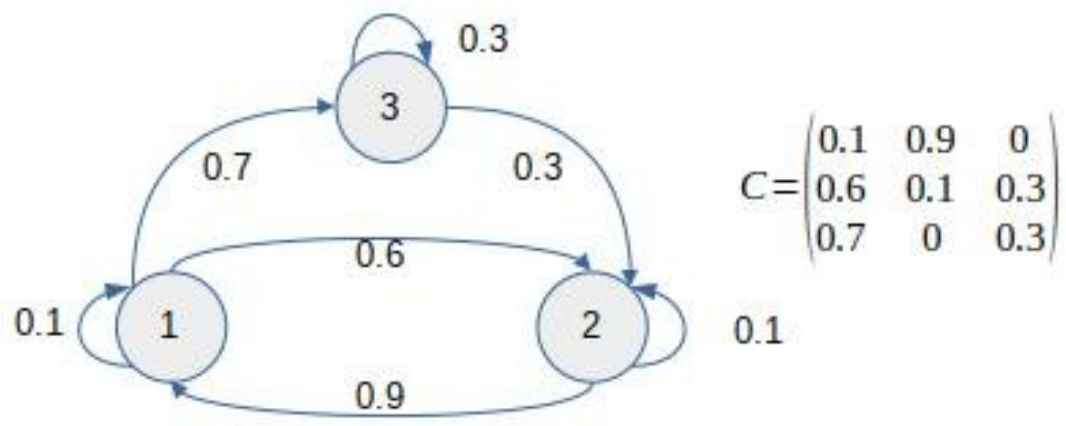

**Figure 4. Social Topology of Individuals**

Intuitively, one may expect that the iterative process of averaging opinions will bring newly formed opinions closer to each other until they reach a consensus. In the following discussion we will explain that the dynamics of opinion formation can be much more complex than one would intuitively expect. A major reason is that the weights put on the opinions of others is not fixed, but time-variant, critically depending on the *n*-agent states. Moreover, $\boldsymbol{C}=[c_{ij}]_{n\times n}$ cannot be formalized as an explicit function of time *t*, but an explicit function of *n*-agent states. The opinion dynamics of *n*-agents thus becomes nonlinear in principle. To simplify the following discussion, it is necessary to first extract the time-invariant component from $\boldsymbol{C}=[c_{ij}]_{n\times n}$ , and it basically represents the stable social relationship of *n*-agents as denoted by $s_{ij}$ . The parameter $c_{ij}$ is thus rewritten by

$$c_{ij}=\frac{\varphi_{ij}\, s_{ij}}{\sum_k \varphi_{ik}\, s_{ik}}$$

$$\varphi_{ij} = 1 \text{ if individual } i \text{ has access to acquire individual } j\text{'s opinion}$$

$$\varphi_{ij} = 0 \text{ if individual } i \text{ has no access to acquire individual } j\text{'s opinion} \qquad (6)$$

In Equation (6) $\varphi_{ij}$ is a kind of binary variables updated based on an agent's state (e.g, positions), describing if agent *i* is socially connected from agent *j* in the topology (e.g., observing agent *j*'s choice or talking to agent *j* to exchange opinions). In contrast $s_{ij}$ is a quantitative measure of social relationship, and it is a real number to indicate to what extent agent *i*'s opinion is possibly impacted by agent *j*. It is obvious that parameter $c_{ij}\geq 0$ is automatically normalized by using Equation (6), and it is not required to normalize $s_{ij} \geq 0$ in order to use Equation (6). Furthermore, in this article we only consider $s_{ij} \geq 0$, implying that agents are milling with each other to form opinions, but not antagonistic to each other ($s_{ij} < 0$). To be consistent with our previous discussion, we strictly let $s_{ii} > 0$, meaning that an agent must support his own opinion to some extent, but will not completely follow others' opinion. To make the things not too complicated $s_{ij}$ is assumed to be constant such that it does not change through the simulation process. In other words, the social weight of two agents is defined as a constant $s_{ii}$ , but whether it is counted in matrix $\boldsymbol{C}=[c_{ij}]_{n\times n}$ entirely depends on boolean parameter $\varphi_{ij}$.

Now an interesting issue is about how $\varphi_{ij}$ is dynamically changing based on agent states. According to the Vicsek flocking model a widely-accepted criterion is that $\varphi_{ij}=1$ when the physical distance $d_{ij}$ is less than $R_i$ , namely, $d_{ij}<R_i$ . In other words, if agent *j* move into a circle of radius $R_i$ surrounding the agent *i*'s position, the opinion of agent *i* will be impacted by agent *j*. In particular, the Vicsek flocking model assumes that each individual agent is given the same radius $R_i$ , and thus the opinion of two agents are influenced mutually. This is a reasonable and relatively straightforward assumption such that group members exchange opinions when they are sufficiently close to each other. As a result, the phase transition occurs when the agent density exceeds a certain threshold in Viscek flocking model. In a similar analysis it is also inferred that the agent opinions as defined by the probability distribution on binary choices will also converge if the agent density exceeds a certain threshold. As below we illustrate numerical results for binary choice process where a simple structural layout is used as shown Figure 2, and consensus of probability distribution is shown in Figure 5.

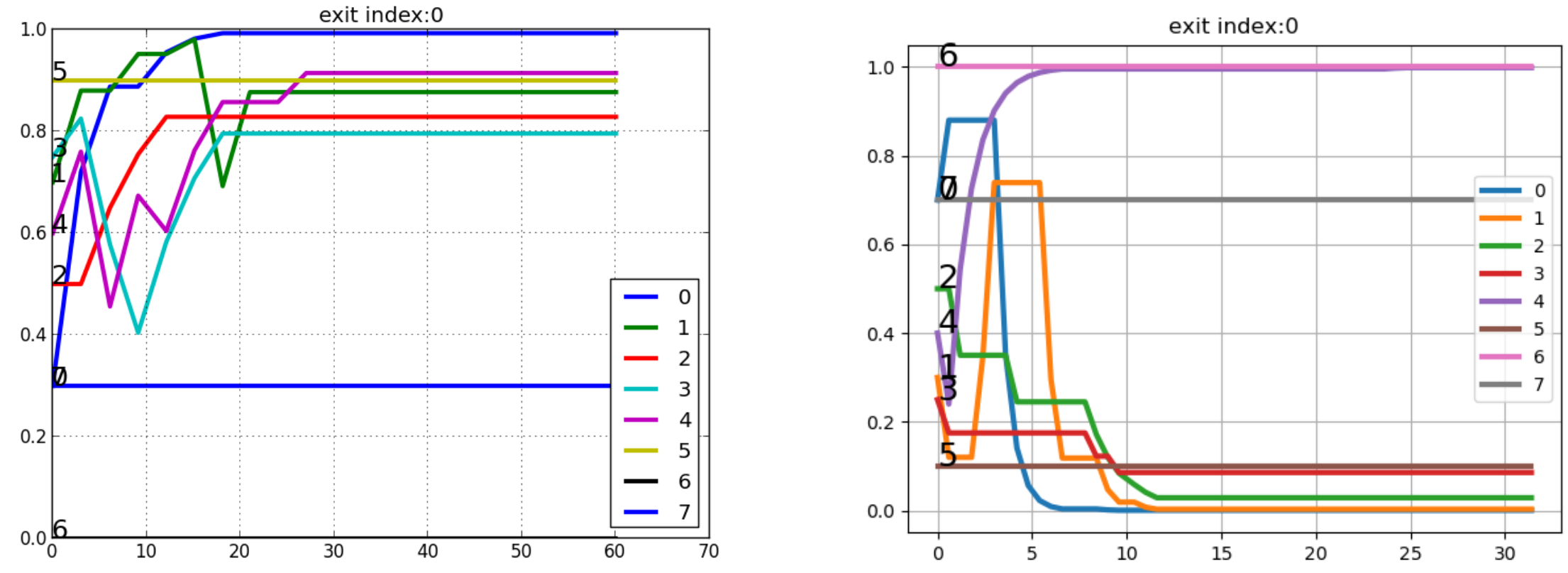

**Figure 5. Simulation results of binary choice process: Consensus or Clustering Effect.**

Usually no obstructions are considered in the Vicsek flocking model and agents are moving freely in an open planar space, just like a flock of birds flying in sky, and thus it is reasonable to use the criterion of $d_{ij}<R_i$ to determine $\varphi_{ij}$=1. In our simulation platform socialArray and crowdEgress (Wang et. al., 2020), however, agents are moving in a multi-compartment layout which consists of obstructions and passageways. Thus, a more complicated criterion is developed based on several lists calculated for each agent, and they are called seeing list and attention list, which describe whether an agent is within the visual field of another or whether an agent pays selective attention to another. We can also use boolean matrices to describe such lists, for example by $\boldsymbol{SEE}=[see_{ij}]_{nxn}$ and $\boldsymbol{ATT}=[att_{ij}]_{nxn}$. The element of $see_{ij}$ =1 means that agent $i$ is able to see agent $j$, and $att_{ij}$ =1 means that agent $i$ pays attention to agent $j$. Obviously, we have $see_{ij}$ =1 if $att_{ij}$ =1, namely an agent can only pay attention to another who is within his or her visual field. Furthermore, the value of $att_{ij}$ is determined by social relationship among agents such that an agent will pay selective attention to those who are in close social relationship. In this article we will not discuss the criteria in detail on how to update $\boldsymbol{SEE}=[see_{ij}]_{nxn}$ and $\boldsymbol{ATT}=[att_{ij}]_{nxn}$. In practical computing these boolean matrices are mainly used to update the value of $\varphi_{ij}$ dynamically in the simulation process.

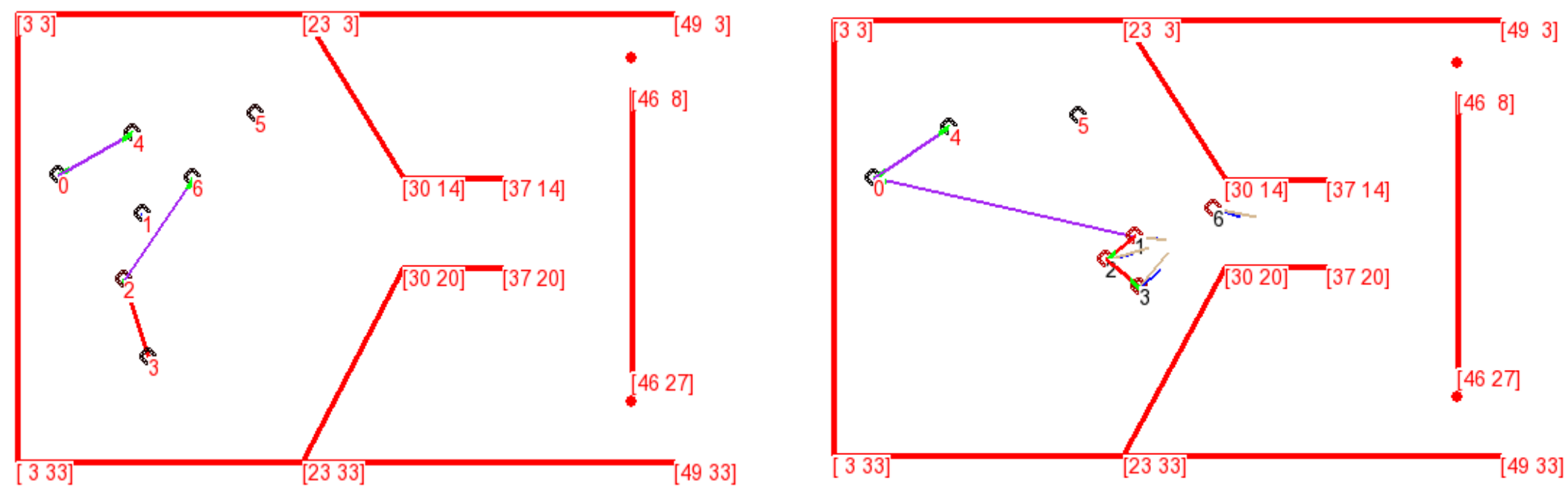

**Figure 5. Time-Varying Social Topology of Individuals: The Social Topology is derived from Matrix $C=[c_{ij}]_{nxn}$ which is a quantitative measurement of social connection among individuals.**

Based on existing model in opinion dynamics, there are several models which are applicable to improve the basic criterion of $d_{ij}<R_i$. For example, an existing theory suggests that interactions bring opinion closer to each other if they are already close sufficiently, and thus an agent tends to selectively follow those with similar opinions, and this algorithm could generate several small groups rather than one large group of consensus. Such a nonlinear opinion dynamics is based on a kind of "cohesive force" between similar opinions. In the following discussion we will follow the basic criterion of $d_{ij}<R_i$, but we will consider the attraction effect by force-based interaction. In particular the social relationship among agents is especially taken into account such that agents in close relationship are attracted and clustered in physical positions, increasing the probability of satisfying the criterion $d_{ij}<R_i$.

## IV. COHESIVE EFFECT AND GROUP SOCIAL FORCE

Based on the opinion dynamics as mathematically formulated in Section 3, we know that individual-level opinions converge towards a common probability distribution. However, even if their distributions on different choices merge together, their decision drawn from such distribution could still be diversified. In other word, when many individuals make decision based on a common converged distribution — for example, [0.5, 0.5] as in a typical binary choice problem, the entire group is still possible to be split into two subgroups — for example, half of them choose A while others choose B. Thus, converged opinion distribution does not mean converged decision and behavior in the physics world as shown in Figure 3. In this process de-individualization effect is necessary to keep the entire group together in collective decision making, and force-based interaction plays a major role as to be discussed in the following two sections.

In order to describe de-individualization effect in our forced based model, there are two necessirites: (i) cohesive force; (ii) self-repulsion. We will talk about these two issues sequentially as follows.

Most of the existing opinion dynamics models including bounded confidence model (Deffuant et al. 2000; Hegselmann and Krause 2002), relative agreement model (Deffuant et al. 2002; Amblard and Deffuant 2004) are based on a sort of "cohesive force" between similar opinions. In this section such cohesion effect will be characterized by using a force-based method, where the group social force is introduced to combine agents into social groups. Such cohesive force-based interaction is necessary for modeling de-individualization effect, and it is not formulated based on similar opinions of agents, but essentially on social relationship among them. The social relationship will be also structured in a matrix-based method. Recall Equation (2) as introduced in Brownian agent in Section 2, the group social force is integrated as below to describe interaction of agents.

$$m_i \frac{d\,\boldsymbol{v}_i(t)}{dt} = \boldsymbol{f}_i^{drv} + \sum_{j(\neq i)} \boldsymbol{f}_{ij}^{soc} \tag{7}$$

Different from the driving force $\boldsymbol{f}_i^{drv}=\boldsymbol{F}^{drv}(\boldsymbol{v}_i^0-\boldsymbol{v}_i)$ as introduced in Section 2, the social-force $\boldsymbol{f}_{ij}^{soc}$ characterizes the social-psychological tendency of agents to keep proper distance with each other in collective motion, and it can be generally denoted by $\boldsymbol{f}_{ij}^{soc}=\boldsymbol{F}^{soc}(d_{ij}^0-d_{ij})$, where $d_{ij}^0$ represents the equilibrium position when agent *i* interacts with agent *j*. A novel form of such interaction force is presented in Wang 2016 and Wang et. al., 2022 by renewing the basic concept of social force in Helbing and Molar, 1995 and Helbing et. al., 2002, and it is mathematically expressed by an exponential form in Equation (8). Similar to the driving force and desired velocity $\boldsymbol{v}_i^0$ as given by Equation (2), the desired interpersonal distance is denoted by $d_{ij}^0$ and it is the target distance in one's mind, specifying the distance that one expects to adapt oneself with others. The physical distance $d_{ij}$ is the distance achieved in the reality.

$$\boldsymbol{f}_{ij}^{soc} = \frac{A_{ij}}{B_{ij}}(d_{ij}^0-d_{ij})\exp\left[\frac{(d_{ij}^0-d_{ij})}{B_{ij}}\right]\boldsymbol{n}_{ij} \quad \text{or} \quad \boldsymbol{f}_{ij}^{soc} = \left(\lambda_i+(1-\lambda_i)\frac{1+\cos\varphi_{ij}}{2}\right)\frac{A_{ij}}{B_{ij}}(d_{ij}^0-d_{ij})\exp\left[\frac{(d_{ij}^0-d_{ij})}{B_{ij}}\right]\boldsymbol{n}_{ij} \tag{8}$$

Here $A_i$ and $B_i$ are positive constants, which affect the strength and effective range about how two agents are interacting to each other. The distance between agent *i* and *j* is denoted by $d_{ij}$, and the sum of the radii of individual *i* and *j* is denoted by $r_{ij}= r_i + r_j$, and $\boldsymbol{n}_{ij}$ is the normalized vector directing from agent *j* to *i.* The geometric features of two agents are illustrated in Figure 6. When the force is applied to isotropic particles or agents, the first mathematical equation in (8) is applied. When the model is used to describe collective motion of living bodies such as human pedestrian or bird flocking, an anisotropic formula of the force is widely applied where the equation is scaled by a function of $\lambda_i$. The angle $\varphi_{ij}$ is the angle between the direction of the motion of agent *i* and the direction to agent *j*, which is exerting the repulsive force on agent *i*. If $0 < \lambda_i < 1$, the force is larger in front of an agent than behind. This anisotropic formula assumes that agents are only able to perceive things in front of them due to limited visual field. Other anisotropic formula can also be applied in pedestrian motion (Chraibi et. al., 2011), and they are effective in a similar way that the force is larger in front of a pedestrian than behind.

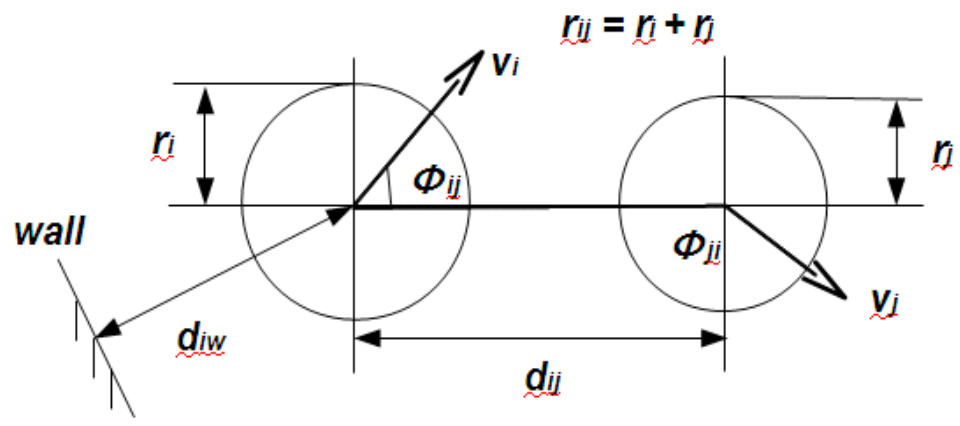

**Figure 6. A Schematic View of Two Agents (See Equation 8)**

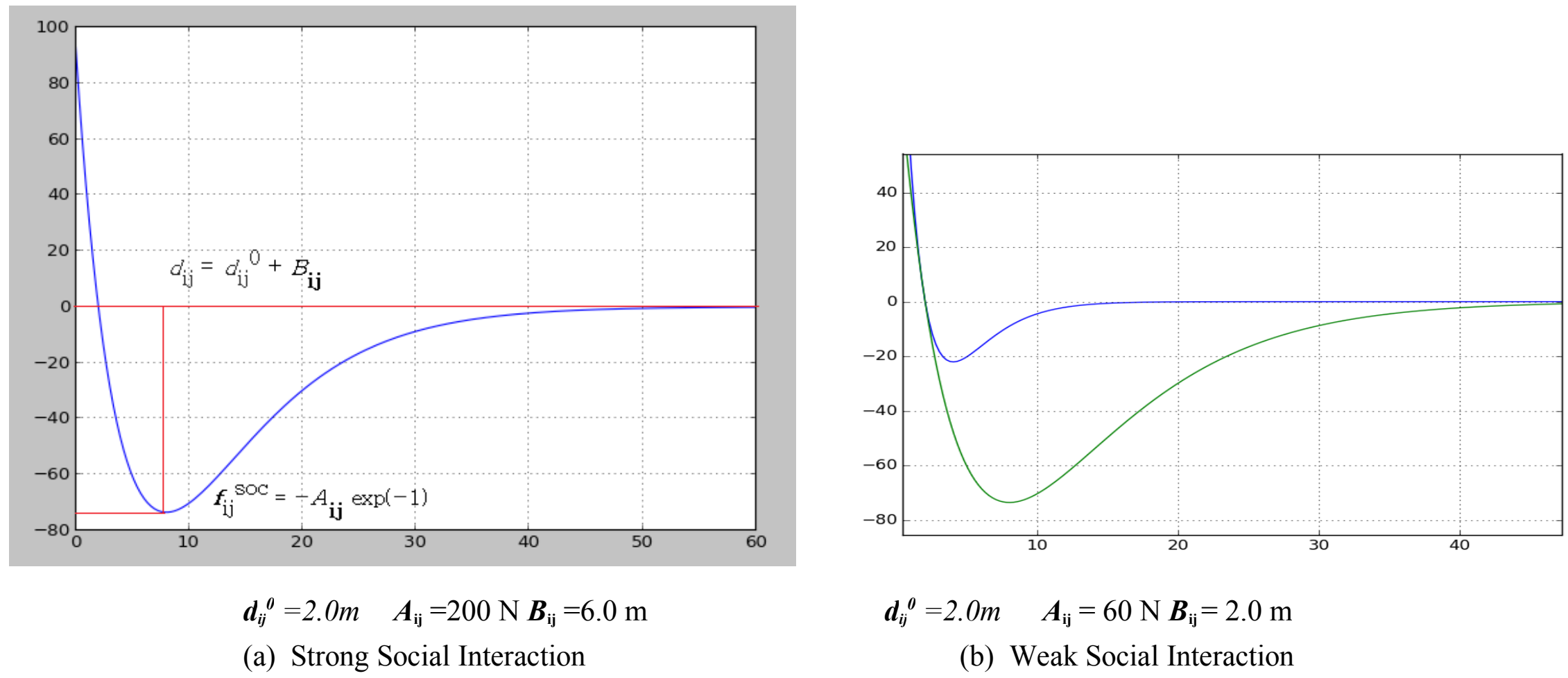

$d_{ij}^{0}$ =2.0m $A_{ij}$ =200 N $B_{ij}$ =6.0 m

(a) Strong Social Interaction

$d_{ij}^{0}$ =2.0m $A_{ij}$ = 60 N $B_{ij}$ = 2.0 m

(b) Weak Social Interaction

**Figure 7. Group social force from individual *j* to individual *i* (*non-anisotropic formula*): (a) To characterize two individuals who know each other, the force includes a negative segment representing attraction as well as a positive segment representing repulsion; (b) When two individuals are strangers, attraction significantly decreases in both strength and the effective range.**

Basically, Equation (8) describes a kind of cohesive effect among interacting agents, and in this article we call it group social force because the mathematical expression of Equation (8) is largely developed based on the traditional social force (Helbing and Molar, 1995 and Helbing, Farkas and Vicsek, 2000). Both of them are in exponential form. The group social force essentially describes attraction and repulsion within the same social context: repulsion makes agents to keep away from each other while attraction aggregates them into social groups. The resulting interaction force is either repulsive or attractive, describing how individuals are self-organized into pattern of collective motion. The mathematical characteristics of Equation (8) is briefly discussed as below. When $d_{ij}$ is sufficiently larger than $d_{ij}^{0}$, the group social force tends to be zero such that agent *i* and *j* have almost no interaction. This is the same as the traditional social force. When $d_{ij}$ is reducing towards $d_{ij}^{0}$, interaction of agent *i* and *j* becomes more and more effective. In particular, the group social force is attraction if $d_{ij}^{0} < d_{ij}$, whereas it becomes repulsion if $d_{ij}^{0} > d_{ij}$. The equilibrium position also exists in $d_{ij}^{0} = d_{ij}$ such that the interaction is also zero. The attraction reaches the extreme value when $d_{ij} = d_{ij}^{0} + B_{ij}$, and the extreme value is given by $\boldsymbol{f}_{ij}^{soc} = -A_{ij}\exp(-1)$. Obviously, increasing or decreasing desired distance $d_{ij}^{0}$ makes the curve move horizontally, and the equilibrium position also shifts correspondingly while the curve shape is not changed. The curve shape is basically determined by parameter $A_{ij}$ and $B_{ij}$. Parameter $A_{ij}$ is a linear scaling factor which affects the strength of the force whereas $B_{ij}$ determines the effective range of the force.

Two plots of Equation (8) are comparatively illustrated in Figure 7(b): The green curve shows that individual *i* is attracted by individual *j* when they are sufficiently close in physical positions, and this scenario suggests that individual *i* is socially tied to individual *j*, probably by certain social relationship such as mother-and-child relationship in sheep herd or human crowd. The comparative curve in blue line does not show such strong social connection because their interaction range and magnitude are both reduced remarkably. In Figure 7 the negative segments of the curves are attraction whereas the positive segments are repulsion (See Equation 8). The effect of repulsion is similar to the traditional social force, and it keeps agents to stay away with each other with proper distance. However, the group social force is different from the traditional social force mainly in the two aspects.

First of all, the desired distance $d_{ij}^{0}$ in the group social force is commonly larger than $r_{ij} = r_i + r_j$ in the traditional formula of social force, where $r_{ij} = r_i + r_j$ represents the sum of physical size of two agents. As mentioned before, the traditional social force is usually considered as short-range interaction (e.g., collision avoidance), and it is effective only when two agents are very close to each other. Thus, it is mainly applied to describing high-density crowd, herd or school. For example, when many individuals compete to pass through a narrow doorway, they become very close to each other and traditional social force is mainly applied in this scenario (Helbing, Farkas and Vicsek, 2000, Hebling et., al., 2002). The group social force is different in a sense that it is relatively a long-range interaction where the desired distance $d_{ij}^0$ is commonly larger than $r_{ij}$, and parameter $B_{ij}$ of group social force is often selected larger than $B_i$ in the traditional social force. In our numerical testing, it is found that $B_{ij}$ is usually in the range of $10^1 \sim 10^{-1}$ while $A_{ij}$ is commonly in the range of $10^2 \sim 10^0$. This issue will be further discussed in detail in numerical testing results.

Secondly and very interestingly, the group social force differs from the traditional social force in a sense that it essentially functions in a feedback manner to make the realistic distance $d_{ij}$ approaching towards the desired distance $d_{ij}^0$. Here $d_{ij}^0$ is the equilibrium position, and this feature is similar to molecular interactions, such as Lennard-Jones potential or Morse potential in molecular dynamics. Moreover, compared with the driving force and desired velocity $\boldsymbol{v}_i^0$ in Equation (2), the desired distance $d_{ij}^0$ is the target distance in one's mind, specifying the distance that one desire to adapt oneself with others. The physical distance $d_{ij}$ is the distance achieved in the reality. The gap between $d_{ij}^0$ and $d_{ij}$ implies the difference between the subjective opinion in one's mind and objective feature in the reality. A difference is that $\boldsymbol{v}_i^0$ and $\boldsymbol{v}_i$ are vectors while $d_{ij}^0$ and $d_{ij}$ are scalars. The general idea is illustrated as below.

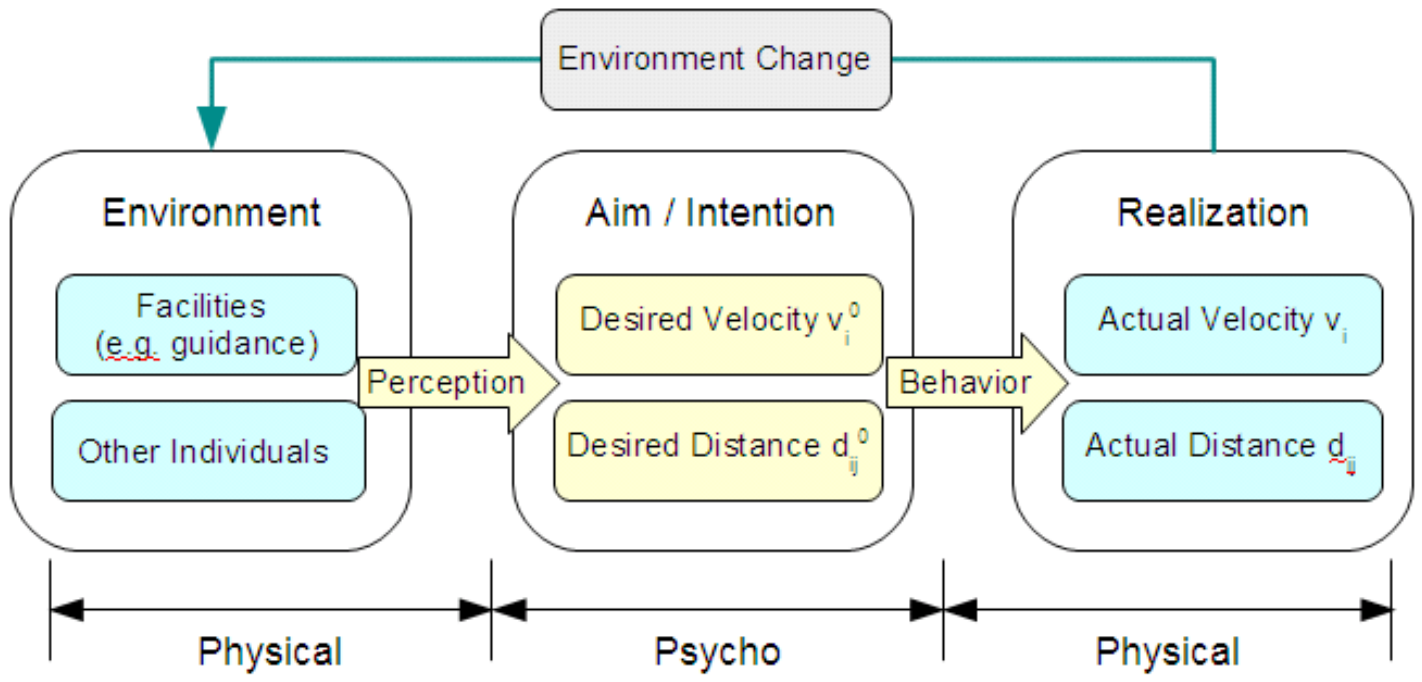

**Figure 8. Perception and Behavior in a Feedback Mechanism**

Furthermore, if we have $d_{ij}^{gap}(t) = d_{ij}^0(t) - d_{ij}(t)$ as an elementary term, it is feasible to integrate the differential and integral of this elementary term into the force, and the force is generalized as below, and this is similar to constructing a PID controller that guide one's behavior toward the target in opinion. The anisotropic formula is not explicitly included in Equation (9), but it could be counted in practical computing process.

$$\boldsymbol{f}_{ij}^{soc} = \boldsymbol{n}_{ij}\left(k_1 \int d_{ij}^{gap}\,dt + k_2 d_{ij}^{gap} + k_3 \frac{d\,d_{ij}^{gap}}{dt}\right) = \boldsymbol{n}_{ij}\left(k_1 \int \left(d_{ij}^0(t) - d_{ij}(t)\right)dt + k_2\left(d_{ij}^0(t) - d_{ij}(t)\right) + k_3 \frac{d\left(d_{ij}^0(t) - d_{ij}(t)\right)}{dt}\right) \quad (9)$$

One issue to be emphasized is the differential term in Equation (9) that leads to a force component of relative velocity $\boldsymbol{v}_{ji} = \boldsymbol{v}_j - \boldsymbol{v}_i$. In fact the force component of has been highlighted in many existing research publications because it helps to offset the possible oscillation phenomenon caused by the interaction force. From the perspective of control theory, the differential term is a widely-used method to offset oscillation, and it corresponds to the relative velocity $\boldsymbol{v}_{ji} = \boldsymbol{v}_j - \boldsymbol{v}_i$, namely the relative velocity of moving individual *j* towards individual *i*. In the following equation if we assume that individual *i* and *j* are located at $\boldsymbol{r}_i$ and $\boldsymbol{r}_j$, and it yields $d_{ij} = |\boldsymbol{r}_i - \boldsymbol{r}_j|$. Here the anisotropic term is omitted and $d^0_{ij}$ is assumed to be constant or it changes relatively much slower than $d_{ij}$, and thus its effect is also omitted. As a result, the third term in the right side of Equation (8) approximates a force component of $\boldsymbol{v}_{ji}$.

$$\frac{d(\boldsymbol{r}_j - \boldsymbol{r}_i)}{dt} = \boldsymbol{v}_{ji} = \boldsymbol{v}_j - \boldsymbol{v}_i = -\boldsymbol{n}_{ij}\frac{d\,d_{ij}}{dt} \approx \frac{d(d_{ij}^0 - d_{ij})}{dt}\boldsymbol{n}_{ij} \quad (10)$$

This term is a best estimate of the future trend of the gap $d_{ij}^{gap}(t)$ based on its current rate of change. It is sometimes called "anticipatory control," which helps reduce oscillation or avoid collision in agent movement. This derivative term has been widely mentioned in many pedestrian models such as the magnetic force model (Okazaki and Matsushita,1993), generalized centrifugal force model (Chraibi et. al., 2011) and many others. Thus, by tuning a force component which is a function of relative velocity $\boldsymbol{v}_{ji}$, the oscillation and collision phenomenon will be significantly mitigated. This effect has been studied in one-dimension analysis (Kretz, 2015).

## V. Social Groups in Array-Based Structure

Group-level behavior is often created beyond the ken of any single individual, and there has been growing realization in social science that such group-level organizations sometimes emerge spontaneously without any central design. Thus, it is reasonable to study such group phenomena in a bottom-up rather than a top-down manner. Agent-based modeling is thus a useful approach, where each individual is mathematically described as the computational unit, and the global structure of many individuals emerges dynamically from their interactions. In this section we will introduce a set of matrices to describe the overall structure of group-level dynamics and discuss the emerging pattern of collective behavior of many agents.

Let $n$ be the number of agents under consideration, and the social topology of $n$ individuals is firstly described by a $n\times n$ matrix $\boldsymbol{C}$ as mentioned above, and its element $c_{ij}$ is timely updated to describe if individual $i$ are able to perceive or acquire opinion of individual $j$ based on their social relationship. For example $c_{ij}$ becomes non-zero when individual $i$ is able to see or talk to individual $j$. Because $\boldsymbol{C}=[c_{ij}]_{n\times n}$ is the matrix used in opinion dynamic model $OPIN(\text{t+1})=\boldsymbol{C}\cdot OPIN(\text{t})$, we will jointly use it with group social force as mathematically described by another three $n\times n$ matrices: $\boldsymbol{A}=[A_{ij}]_{n\times n}$, $\boldsymbol{B}=[B_{ij}]_{n\times n}$ and $\boldsymbol{D}^0=[d^0_{ij}]_{n\times n}$, respectively (See Equation (8)). Generally speaking, $\boldsymbol{C}$, $\boldsymbol{A}$, $\boldsymbol{B}$, and $\boldsymbol{D}^0$ are asymmetrical and could be time-variant.

$$\boldsymbol{C}=[c_{ij}]_{n\times n} \qquad \boldsymbol{A}=[A_{ij}]_{n\times n} \qquad \boldsymbol{B}=[B_{ij}]_{n\times n} \qquad \boldsymbol{D}^0=[d^0_{ij}]_{n\times n} \tag{10}$$

The group social force is specified by the matrices $\boldsymbol{D}^0$, $\boldsymbol{A}$ and $\boldsymbol{B}$, and the method has been partly tested in FDS+Evac as well as our simulation platform SocialArray and CrowdEgress (Wang et. al., 2024). A testing result is illustrated in Figure 9. where two groups are identified in this scenario. One group consists of individual 1, 2, 3, 6 while another group consists of individual 0 and 5. Both groups are moving towards the passageway. In this scenario individual 0 also pays attention to individual 1, and thus two groups also have a kind of connection at that moment. In a sense social group is not a stationary concept in our agent-based modeling framework, but a time-variant feature that critically depends on the agent state. Thus, in this example it is also possible for two groups to merge into one large group in certain conditions. In other words, as we discussed in Section 3, the social topology of group members change dynamically, resulting in a self-organized phenomenon during the collective movement of many individuals.

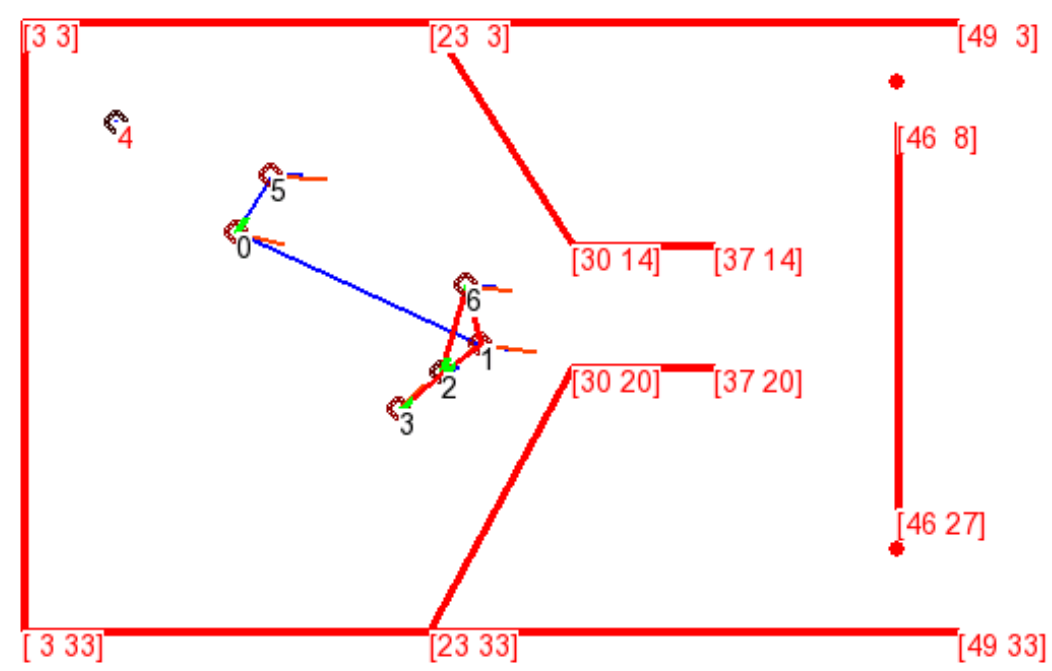

**Figure 9. Simulation of Group Dynamics in CrowdEgress.**

Recall Equation (8), and a typical pattern is described as the leader-and-follower group, where motion of certain agents are mainly motivated by themselves, and if others would like to follow them, they become leaders in the group. Thus, if individual $i$ is the leader in a group, his or her motion is mainly motivated by the self-driving force based on desired velocity. In contrast a follower is mainly motivated by surrounding others. In other words, if individual $i$ is the follower in a group, his or her motion is primarily governed by group social force, and the self-driving force is secondary. As mentioned before, imbalance (asymmetry) of $\boldsymbol{f}_{ij}^{soc}$ and $\boldsymbol{f}_{ji}^{soc}$ will contribute to model leadership in crowd behavior. As a result, the leader will attract his surrounding people, but not easily be attracted by them. In brief an individual's motion can be classified into two types. One type of motion is primarily motivated by the self-driving force and is called active motion. The other type of motion is largely motivated by surrounding people, and is called passive motion. In general, an individual's motion is a combination of both types, but we can differentiate such two types in simulation and identify whether one's motion is either active or passive. As shown in Figure 9 an individual in active motion often moves in the front of a group, and thus individual 1 seems to lead the group in Figure 9. Individuals in passive motion usually move behind, such as individual 2, 3 and 6.

Now an interesting issue is that we suppose $d_{ij}^0$ and $d_{ji}^0$ also exist on the opinion scale. Thus, the matrix-based opinion model could be also applied as $\boldsymbol{D}^0(t+1)=\boldsymbol{C}\cdot\boldsymbol{D}^0(t)$, where $d_{ij}^0$ is the desired distance that agent *i* desires to keep to agent *j*, and it is dynamically adjusted as a linear combination of all other agents' desired distances to agent j. In other words, we have $d_{ij}^0(t+1)=\Sigma_k\, c_{ik}\cdot d_{kj}^0(t)$ to update each element in $\boldsymbol{D}^0$. In the similar manner we also have $\boldsymbol{A}(t+1)=\boldsymbol{C}\cdot\boldsymbol{A}(t)$ and $\boldsymbol{B}(t+1)=\boldsymbol{C}\cdot\boldsymbol{B}(t)$ to further tune the group social force in a dynamic process. Here matrix $\boldsymbol{C}=[c_{ij}]_{nxn}$ is an explicit function of *n*-agent states, and opinion dynamics of *n*-agents thus becomes nonlinear. One issue to be mentioned is an interesting variation of the model by using the random gossip algorithm. In this algorithm agent *i* only randomly select one agent in social connection for interaction at each time step, not interacting with all the agents in social connection (Boyd et. al., 2006), and it is easy to achieve a kind of balance between $d_{ij}^0$ and $d_{ji}^0$ by using this algorithm (Wang, 2016).

With combination of group social force and opinion dynamics a kind of convergent pattern is supposed to emerge in a crowd. Here the social groups and opinion dynamics are related but different concepts. Social groups emphasizes whether there is a social tie between individuals, and such a social tie aggregates individuals into a group. The opinion dynamics, or generally considered as herding effect, emphasizes how an individual's opinion interacts with others' to form a common motive in collective motion. You may meet your friend on the street, but if you do not have a common destination, you and your friend head to each destination individually after greeting or talking briefly. Another example is evacuation of a stadium where people follow the crowd flow to move to an exit. There are a multitude of small groups composed of friends or family members, and they keep together because of their social ties. These small groups also compose a large group of evacuees, and herding behavior widely exists among these small groups, contributing to form a collective pattern of motion.

In sum, the group social force makes individuals socially bonded with each other in physical positions, and opinion dynamics further describes how an individual bring others' opinion in forming his or her own opinion, and thus help to form a common motive in collective motion. In a sense we think that opinion dynamics is a more fundamental feature to describe how individuals interact in social context. In practical computing, it is suitable to first give matrix $\boldsymbol{C}$ as a quantitative measure of social relationship, and matrices $\boldsymbol{A}$, $\boldsymbol{B}$, $\boldsymbol{D}^0$ are next specified in consistency. In brief, because social relationship is described by both of opinion dynamics and group social force, such two features are basically inter-dependent, and it is important to define them in consistency such that matrices $\boldsymbol{A}$, $\boldsymbol{B}$, $\boldsymbol{D}^0$ should be given in consistency with matrix $\boldsymbol{C}$. In practical computing, we usually first define matrix $\boldsymbol{C}$ for quantitative measurement for social relationship of *n* agents, and the matrices $\boldsymbol{A}$, $\boldsymbol{B}$, $\boldsymbol{D}^0$ are next deduced in consistency with matrix $\boldsymbol{C}$. In this paper we present a simple rule to specify $\boldsymbol{A}$, $\boldsymbol{B}$, $\boldsymbol{D}^0$ in consistency with matrix $\boldsymbol{C}$ as below.

$$\begin{aligned} &\text{if } c_{ij}\neq 0, \quad A_{ij}=100c_{ij}, \quad B_{ij}=0.001+10c_{ij}, \quad d_{ij}^0=1.0 \\ &\text{if } c_{ij}=0, \quad A_{ij}=0 \Rightarrow \boldsymbol{f}_{ij}^{soc}=0 \end{aligned} \tag{11}$$

The rule of (11) implies that both intensity and effective range of group social force increases with $c_{ij}$ while the equilibrium position $d_{ij}^0$ remains constant for any two agents socially linked in the topology, i.e., $c_{ij}\neq 0$. As for any two agents not socially linked in the topology, i.e., $c_{ij}=0$, the group social force is zero. The above algorithm has been well tested in our simulation platform CrowdEgress [10] and SocialArray.

Now we will summarize the above computational models of self-propelled agents, which build social structures in a matrix-based approach. In a general sense each agent is mathematically formulated with both opinion, decision and action/motion, and their opinions generate motive of action such that agents interact and move in a 2-dimensional space. Very importantly, interaction of individual agents occur at levels of both opinion models and forced-based actions. In an individual sense, motion is self-propelled by agents' opinion. When they are aggregated into groups in collective behavior, they become a complex interactive system, and an important issue to be addressed is how their group behavior are aggregated in a deterministic manner if their individualistic opinions are diversified to some extent in a probabilistic sense.

To better explain this problem, we will recall the binary choice example as illustrated in Figure 2 and 5. Suppose an individual is facing a decision of choosing either A or B. Such two choices could co-exist in opinion in a probabilistic sense. For example, you may prefer option A with 0.7 probability, and option B with 0.3 probability. However, when such an opinion is realized into behavior in the physical world, a decision should be reached by selecting one option definitely. Things become complex if many individuals interact in order to reach a collective decision. Suppose you and your friends decide together either eating at home or going to a restaurant, each one may have his or her indi-vidual preference in mind, as described by a probability distribution. When such probabilistic measurement is realized into behavior in a deterministic sense, a key problem that we are interested in is whether an individual will comply his or her individual choice with the group-level choice, especially when such two choices are not the same. In other words, our agent-based model assumes that it is not free for an individual to join a social group, the trade-off is de-individualization of oneself, and this phenomenon is consistent with existing socio-psychological principles (Fromm, 1941). Thus, within a social group the group members may not all agree to select a common target after exchanging the information, and the group dynamics is formulated in a sense that someone may lose part of his or her individual motive in order to join the group. In the history of psychology study this effect was initially described in LeBon's book, "The Crowd: A Study of the Popular Mind"(LeBon, 1895). In this book the author described a phenomenon that an individual seems forgetting the original motive when immersing oneself in the crowd, and thus the individual-level motive is completely replaced by the collective motive of crowd. This is usually called de-individualizing process in psychological studies.

Very interestingly, we notice that the diagonal elements in matrix $\boldsymbol{D^0}$, $\boldsymbol{A}$ and $\boldsymbol{B}$ imply a kind of force to oneself, where $d_{ii}$ = 0. However, Equation (8) implies the force is zero because $\boldsymbol{n}_{ii}$ is a zero vector. Thus, we may modify the vector slightly such that an individual can implement a kind of force to oneself, that is, $\boldsymbol{n}_{ii}$ is replaced by a normalized vector with the opposite direction of the driving force $\boldsymbol{f}_i^{drv}$, and this force $\boldsymbol{f}_{ii}^{soc}$ is called self-repulsion in this paper, which means a kind of social force exerted to oneself.

$$\boldsymbol{f}_{\boldsymbol{ii}}^{\boldsymbol{soc}} = A_{ii}(d_{ii}^0 - d_{ii})\exp\left[\frac{d_{ii}^0 - d_{ii}}{B_{ii}}\right]\boldsymbol{n}_{\boldsymbol{ii}} = A_{ii}d_{ii}^0\exp\left[\frac{d_{ii}^0}{B_{ii}}\right]\boldsymbol{n}_{\boldsymbol{ii}} \quad \longrightarrow \quad \boldsymbol{f}_{\boldsymbol{ii}}^{\boldsymbol{soc}} = A_{ii}d_{ii}^0\exp\left[\frac{d_{ii}^0}{B_{ii}}\right](-normalize(\boldsymbol{f}_{\boldsymbol{i}}^{\boldsymbol{drv}})) \tag{12}$$

In a socio-psychological sense the self-driving force is formed by conscious mind of an individual, and it generates one's motivation of behavior. The self-repulsion refers to the unconscious mind, and it may be against the conscious motive that we are aware of. Thus, Equation (12) assumes that direction of self-repulsion is opposite to the driving force. This model is meaningful to understand certain crowd behavior, for example, the "automaton conformity" as depicted in Fromm, 1941. That is, when an individual intentionally immerses oneself in the crowd, he or she may lose part of his individual feature such as his original motivation, and thus simply follow the collective motive of the crowd in order to gain a sense of safety. In another sense, we may explain such "automaton conformity" as a special kind of herding effect occurring in human society. Thus, the force specified by Equation (12) is useful to neutralize the effect of self-driving force $\boldsymbol{f}_i^{drv}$. In this situation the group social force will play an important role and it become dominant such that an individual's motive is replaced by the crowd motive, especially by the leader's motive in the crowd.

However, a major difficulty is how to properly adjust the value of $A_{ii}$, $B_{ii}$ and $d_{ii}^0$ to formulate such self-repulsion in consistency with social group behavior. Another useful method is using matrix $\boldsymbol{C}=[c_{ij}]_{nxn}$ to formulate self-repulsion. As mentioned before, $\boldsymbol{C}=[c_{ij}]_{nxn}$ is more fundamental to describe social relationship among individuals. In particular, individualistic behavior is dominant if $c_{ii}$ is high whereas herding behavior dominates if $c_{ii}$ is low, and $c_{ii}$ indicates how an individual keeps balance between one's own opinion and others' opinions, and thus it indicates to what extent one immerses oneself in the group or crowd. The self-repulsion is formulated as below.

$$\boldsymbol{f}_{ii}^{soc}=[1-\exp(-\beta_i)](-\boldsymbol{f}_i^{drv}) \qquad \beta_i = (1-c_{ii}) / c_{ii} \tag{13}$$

Parameter $\beta_i \geq 0$ indicates level of one's immersion in crowd. In a sense $\beta_i=0$ denotes $\boldsymbol{f}_{ii}^{soc}=0$ such that one's conscious mind is independent, not influenced by other people. As $\beta_i$ increases, the self-repulsion $\boldsymbol{f}_{ii}^{soc}$ goes towards $-\boldsymbol{f}_i^{drv}$ so that the individualistic feature is neutralized by $\boldsymbol{f}_{ii}^{soc}$, meaning that people immerse themselves in the crowd. In order to apply Equation (13) in practical computing we need to assume $c_{ii}$ changes dynamically in range of (0, 1]. In other words, the model accepts the case that an individual becomes completely oneself when $c_{ii}$ = 1, such as the leader of a group. The tendency of following others increases as $c_{ii}$ goes towards 0, but we suppose that an agent will not completely lose oneself and thus $c_{ii}$ >0. Furthermore, the above model requires that $c_{ii}$ is a variable, which is to be updated in the computational process. How to update $c_{ii}$ is a little complicated issue, and it refers to more study topic in socio-psychological study, and thus we will not cover this issue in this article.

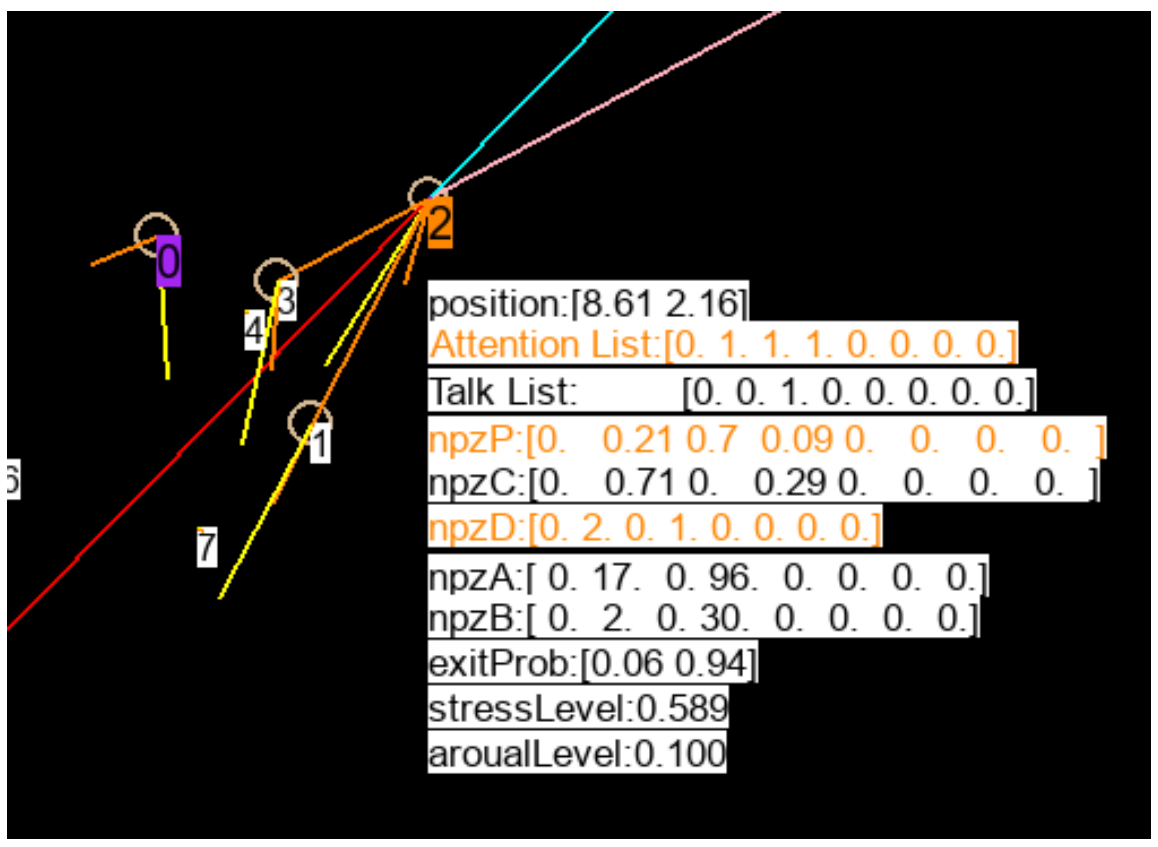

**Figure 10. Illustration of Self-Repulsion in Social Groups**

In Figure 10 we illustrate a simulation scenario of our program socialArray, where the cyan line represents the self-repulsion formulated by Equation (13), and it is implemented on agent 2, which neutralizes the effect of self-driving force $\boldsymbol{f}_i^{drv}$ as represented by the red line in Figure 10. In sum, unconsciousness is an interesting topic in psychological studies, and it is a fantastic issue to be further investigated.